\documentclass[sigconf, nonacm]{acmart}

\newcommand\vldbavailabilityurl{https://gitlab.rlp.net/mvee}
\def\BibTeX{{\rm B\kern-.05em{\sc i\kern-.025em b}\kern-.08em
    T\kern-.1667em\lower.7ex\hbox{E}\kern-.125emX}}

\usepackage{booktabs} % For formal tables
\usepackage{subcaption}
\usepackage{listings}
\usepackage{pythonhighlight}
\usepackage{tikz}
\usepackage{pdfpages}

\lstdefinestyle{pseudo}{
  frame=tb,
  language={c++},
  deletekeywords={with},
  aboveskip=2mm,
  belowskip=2mm,
  captionpos=b,
  showstringspaces=false,
  columns=flexible,
  basicstyle={\footnotesize\ttfamily},
  numbers=left,
  numberstyle=\tiny \color{black},
  keywordstyle=\color{blue},
  commentstyle=\color{magenta},
  frame=none,
  breaklines=true,
  breakatwhitespace=true,
  tabsize=3,
}

\newcommand*\circled[1]{%
  \tikz[baseline=(char.base)]{
    \node[
      shape=circle,
      draw=black,
      fill=black,
      text=white,
      inner sep=0.6pt
    ] (char) {#1};
  }%
}

\newcommand{\smalltt}[1]{{\texttt{\small #1}}}

\newcommand{\m}{MVEE}
\newcommand{\mc}{MVEE core}

\setcopyright{rightsretained}

\acmDOI{}

\acmISBN{XXX}

\begin{document}
\title{The Case for Multi-Version Experimental Evaluation (\m{})}
  
\author{Simon J\"orz}
\affiliation{%
  \institution{Johannes Gutenberg University}
  \city{Mainz}
  \country{Germany}}
\email{joerzsim@uni-mainz.de}

\author{Felix Schuhknecht}
\affiliation{%
  \institution{Johannes Gutenberg University}
  \city{Mainz}
  \country{Germany}}
\email{schuhknecht@uni-mainz.de}

% \renewcommand{\shortauthors}{B. Trovato et al.}

\begin{abstract}
In the database community, we typically evaluate new methods based on experimental results, which we produce by integrating the proposed method along with a set of baselines in a single benchmarking codebase and measuring the individual runtimes. If we are unhappy with the performance of our method, we gradually improve it while repeatedly comparing to the baselines, until we outperform them. 
While this seems like a reasonable approach, it makes one delicate assumption: We assume that across the optimization workflow, there exists only a \textit{single} compiled version of each baseline to compare to. However, we learned the hard way that in practice, even though the \textit{source code} remains untouched, general purpose compilers might still generate highly different \textit{compiled code} across builds, caused by seemingly unrelated changes in other parts of the codebase, leading to flawed comparisons and evaluations.
To tackle this problem, we propose the concept of \textit{Multi-Version Experimental Evaluation (\m{})}. \m{} automatically and transparently analyzes subsequent builds on the assembly code level for occurring ``build anomalies'' and materializes them as new versions of the methods. As a consequence, all observed versions of the respective methods can be included in the experimental evaluation, highly increasing its quality and overall expressiveness.
\end{abstract}

\maketitle

\ifdefempty{\vldbavailabilityurl}{}{
\vspace{.3cm}
\begingroup\small\noindent\raggedright\textbf{Artifact Availability:}\\
The source code, data, and/or other artifacts have been made available at \url{\vldbavailabilityurl}.
\endgroup
}

\section{Introduction and Motivation}

Traditionally, database research heavily relies on experiments. A typical workflow for an evaluation of a new method looks as follows: We first integrate the method along with a baseline into a single codebase. Then, we run both on the same workload and compare the individual runtimes. 
%If the results show that our proposed method is not yet fast enough, we incrementally optimize it. 
If our method is not fast enough, we incrementally try to optimize while re-running the benchmark until we reach our goal.
While at first glance, this looks like a reasonable workflow, it unfortunately relies on the assumption that the incremental builds of the codebase we compare with each other are actually \textit{comparable}. More concretely, we assume that all local modifications we apply to our method also remain local in the actual build and do not affect any other parts of the codebase, such as the generated code of the baseline method. 

\begin{figure}[h!]
  \vspace*{-0.3cm}
  \centering
  \begin{subfigure}[b]{\linewidth}
    \centering
    \includegraphics[page=1, width=\linewidth, trim={0 20cm 22cm 0.5cm}, clip]{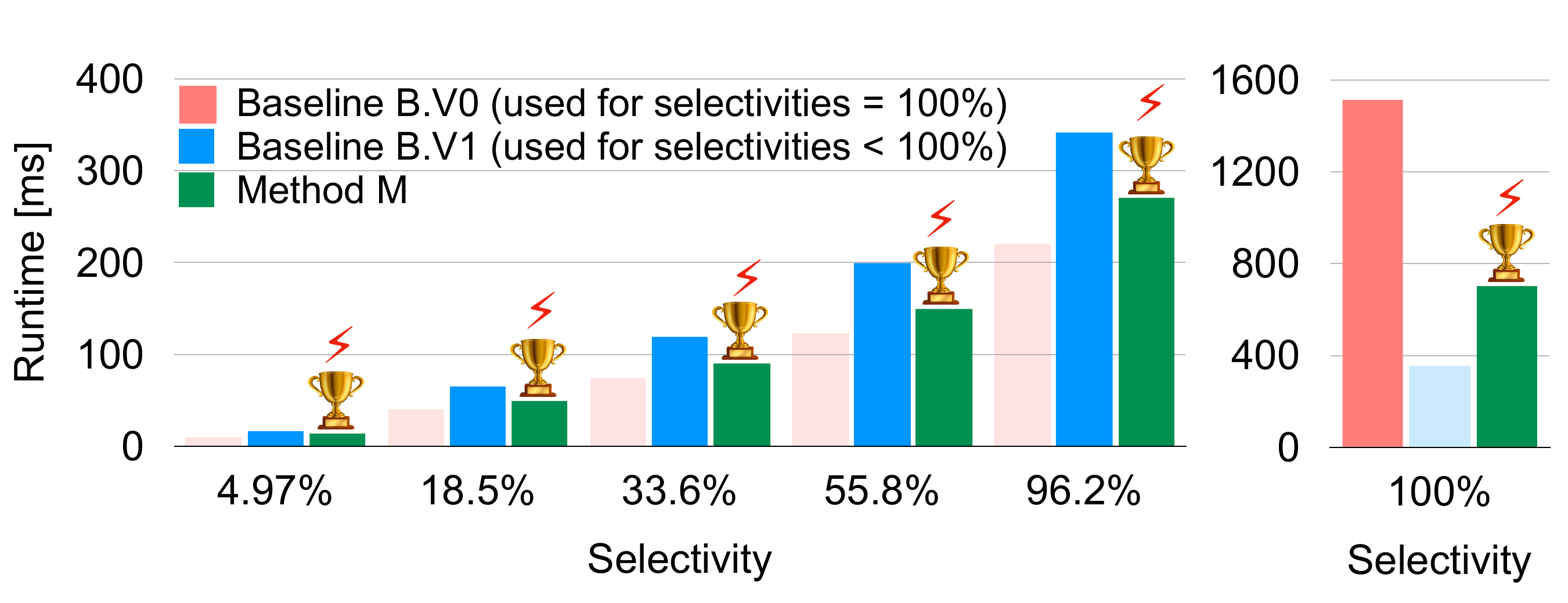}
    \caption{Problem 1: Merging results from different builds $\rightarrow$ false interpretation.}
    \label{fig:motivation:problem1}
  \end{subfigure}
  \begin{subfigure}[b]{\linewidth}
    \centering
    \includegraphics[page=2, width=\linewidth, trim={0 20cm 22cm 0}, clip]{figures/comp_madness_reference.pdf}
    \caption{Problem 2: Showing only the results of one build and ``forget'' about another potentially better build $\rightarrow$ false interpretation.}
    \label{fig:motivation:problem2}
  \end{subfigure}
  \begin{subfigure}[b]{\linewidth}
    \centering
    \includegraphics[page=3, width=\linewidth, trim={0 22.5cm 22cm 0}, clip]{figures/comp_madness_reference.pdf}
    \caption{Solution~\m{}: Actively including the results of \textit{all} seen versions to get the most complete picture $\rightarrow$ meaningful interpretation.}
    \label{fig:motivation:mvee}
  \end{subfigure}
  \vspace*{-0.7cm}
  \caption{Evaluation workflow with and without~\m{}.}
    \vspace*{-0.2cm}
\label{fig:motivation}
\end{figure}

Unfortunately, we learned in previous work~\cite{lit:compilation_anomalies} that this assumption cannot be safely made when using a general-purpose compiler such as \smalltt{gcc}: 
After experiencing unexplainable runtime variances across builds in one of our research projects~\cite{lit:storage_views} that evaluated scan-accelerating index structures across different selectivities, a deep investigation revealed irregularities across these builds on the assembly level: The generated code of a baseline method~\smalltt{B} was heavily affected by certain changes in completely unrelated code parts within the same compilation unit, presumably triggered by one of the many optimization heuristics~\cite{lit:gcc_manual, lit:gcc_opt1, lit:gcc_opt2, lit:gcc_opt3} used within the compiler. This effectively resulted in the unpredictable occurrence of two very different versions~\smalltt{B.V0} and \smalltt{B.V1} of the same baseline. 

If these versions go unnoticed one could easily miss comparing to the ``right'' version. Depending on the workflow, we have to differentiate between the following two cases:
In the first case, results from \textit{different} builds are gathered. Figure~\ref{fig:motivation:problem1} showcases this with real data from~\cite{lit:compilation_anomalies}, where the baseline result from one build is used in the plot for a selectivity of~$100\%$, whereas the results from another build are used for the remaining selectivities. As the example shows, if we are unaware of the existence of the two baseline versions~\smalltt{B.V0} and \smalltt{B.V1}, we will conclude that \smalltt{M} generally outperforms the baseline. In reality, however, we just compared against the versions that perform worse in the respective situations.
In the second case, all results originate from a \textit{single} build, but from an arbitrary one, as showcased in Figure~\ref{fig:motivation:problem2}. While this at least ensures consistency, we might still draw wrong conclusions as we can miss better versions: In our example, this is the case for a selectivity of $100\%$, where we compare \smalltt{M} against the worse version \smalltt{B.V0}, while \smalltt{B.V1} would perform significantly better. 

\vspace*{-0.2cm}
\subsection{Multi-Version Experimental Evaluation}
\label{ssec:mvee_intro}
\vspace*{-0.1cm}

Unfortunately, as long as we deal with complex black-box compilers, we have to assume that such anomalies occur in practice, even if we try to counter-steer the generation of different versions via a stabilization tool~\cite{lit:stabilizer}. 
%While we were able to notice and handle the anomaly in the specific case of~\cite{lit:compilation_anomalies}, we were left with an unpleasant feeling: 
%How often did this problem go unnoticed in experimental evaluations? 
%How often did one method win over another not because it was inherently better, but because their code generation was positively or negatively influenced by seemingly unrelated code?
%Unfortunately, as long as we deal with complex black-box compilers, we have to assume that such anomalies can occur in practice. 
However, instead of simply ignoring the generated versions, we argue to actively incorporate them into the experimental evaluation workflow: If the generated assembly code of a baseline method suddenly changes from one build to the other due to a compilation anomaly without a corresponding change in the source code, we do not ignore this, but treat the anomaly as a new compiled version of the baseline method. At the same time, we do not forget about the old compiled version, but also keep it, along with the experimental results produced therefrom. As a consequence, as shown in Figure~\ref{fig:motivation:mvee}, we can include the results from \textit{all} seen compiled versions to get a complete picture of whether our method \textit{actually} beats the baseline or not.

In the following, we materialize this concept as what we call \textit{Multi-Version Experimental Evaluation (\m{})}. We integrated \m{} as a comfortable extension for the VSCode IDE, which we will showcase in this demonstration.
During the development workflow, 
\m{}~automatically and transparently analyzes the relevant compiled code of generated builds on the \smalltt{x86} assembly level for occurring anomalies.
As soon as an anomaly is detected, \m{} does not only report this to the developer for closer inspection by showing the corresponding assembly and source code portions, but automatically registers the change as a newly seen compiled version of the corresponding source code section, effectively building up a version graph that is shown to the user. Additionally, it stores the experimental results of the corresponding run and maps them to this particular version, which allows to include these results in the produced plots.  

% Figure~\ref{fig:motivation} shows the impact of this on typical optimization workflows: Without \m{} (Figure~\ref{fig:motivation:before}), we might miss an already seen better version of the baseline and stop optimizing our method too early. With \m{} (Figure~\ref{fig:motivation:mvee}), all observed compiled versions are dragged along, resulting in a more complete and meaningful comparison.   

Before we jump into the description of the workflow, let us clearly define what we consider as equivalent builds and what as anomalies. 
We consider two sequences of assembly code lines~$S_1$ and $S_2$ as equivalent if the following three conditions are all met:
(1)~Every instruction in $S_1$ also exists in $S_2$ and vice versa.
(2)~Every instruction in $S_1$ operates on the same data as the corresponding instruction in $S_2$.
(3)~The control flow of $S_1$ is the same as of $S_2$. 
Consequently, any pair of assembly code line sequences that violates our equivalency definition is considered as an anomaly. 
Note that within these checks, we currently make the following relaxations for efficiency: 
First, we ignore indirect jumps based on register content. This allows us to check for these conditions statically by analyzing the code without requiring any run-time data.  
Second, we ignore the register assignment, as anomalies would be detected on the instruction and control flow level already. 
%This relies on the practical assumption, that despite producing performance anomalies, the compiler still generates correct code.  

%\m{} works on assembly level to detect changes in the code assembled by general-purpose compiler.
%It is designed to detect changes at assembly code level that have an impact on
%the run-time, but might not be directly observable to the programmer.
% That is, the run-time behavior changes, while the computational behavior of the program does not change.
% It is not guaranteed whether the code generated by a general-purpose compiler is a ``reasonable''
% implementation of the given source code.
%For the intends and purposes of \m{} it is safe to assume that code generated by a general-purpose compiler
%implements the source code as intended, as compilers are designed and thoroughly tested for this purpose.
% Yet, evidence shows that the code they generate is not always equally efficient.
%\m{} assumes that jumps are equally expensive regardless of the distance. %While this is not the case
%in real world scenarios \todo{is it?}, \m{} is designed to work on small code bases that fit into a single
%source file where jump distances are small. 

% short comings
% - jumps to memory locations or immediates
% - jump tables
% - register renaming
% With this design \m{} currently has a few shortcomings.
% It is hard to predict the data stored at memory locations or in registers with accuracy.
%Thus, \m{} currently gathers no knowledge about run-time data and ignores indirect jumps using registers.

\section{\m{} Workflow}

To demonstrate how \m{} operates, we consider an exemplary development workflow, where we want to
compare the methods~\smalltt{M}, \smalltt{B0}, and \smalltt{B1}. Method~\smalltt{M}~represents our
own method that we optimize incrementally, whereas \smalltt{B0}~and~\smalltt{B1}~represent
baseline methods. 
%As explained before, we consider the situation where all three methods share a common codebase and each run measures the runtime of all methods. 

% \vspace*{-0.2cm}
% \subsection{Bootstrapping}
% \label{ssec:bootstrapping}
% \vspace*{-0.1cm}

To bootstrap the system, the user first communicates to the \m{}~extension which code sections are
actually relevant for the experimental evaluation and should be monitored. Further, the user has to
introduce an identifier for each code section, such that \m{} can later map both active
modifications as well as occurred anomalies in the code to the corresponding sections.  
%Communicating the start and end of code sections reliably is actually harder than it seems since
%function calls or assembly comments used as markers might be moved around or eliminated entirely by
%the compiler. We have to assure that both cannot happen, as the markers are required to be at the
%correct positions in the generated assembly code. 
This is done by packing the code of interest in a run method and surrounding the call to
this method by calls to our mark pre-processor definitions \smalltt{gen\_begin\_mark()} and
\smalltt{gen\_end\_mark()}, as shown in Listing~\ref{listing:markers}. These mark definitions
receive the input, as well as the output of the run method and create a dependency with it.

\begin{lstlisting}[style=pseudo, caption={Marking the code sections to monitor.}, label={listing:markers}, xleftmargin=4.1ex, escapechar=|]
size_t input_B1 = 42;
gen_begin_mark(B1, size_t, input_B1);
size_t res_B1 = run_B1(input_B1); // monitored for anomalies
gen_end_mark(B1, size_t, res_B1);
\end{lstlisting}

\begin{figure}[h!]
    \centering
    \includegraphics[page=3, width=.97\columnwidth, trim={0cm 52cm 14cm 0}, clip]{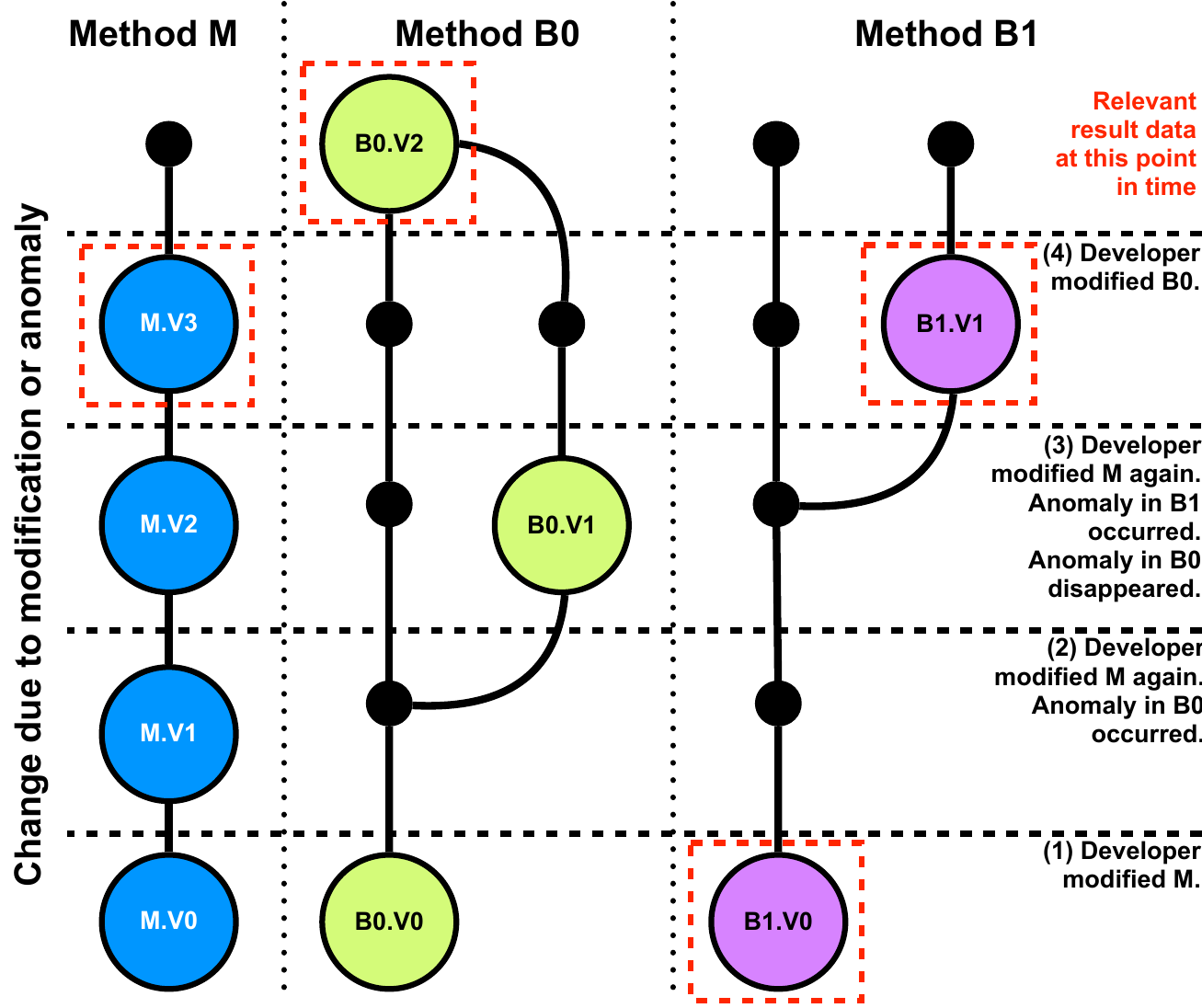}
    \vspace*{-0.2cm}
    \caption{The version graph built up by~\m{}. Each colored circle represents a newly seen compiled version of the method. A source code modification continues respectively merges paths and an anomaly forks the path.}
    \label{fig:version_graph}
    \vspace*{-0.3cm}
\end{figure}

After bootstrapping, let us go through the development workflow for a couple of steps and see how
\m{} handles the occurring effects. Based on the modification and anomaly detection, \m{} builds up
a \textit{version graph} as shown in Figure~\ref{fig:version_graph}. 
This version graph keeps track of all occurred compiled versions for the individual methods.
More importantly it allows to identify at any point in time which versions must currently be
considered for a complete result interpretation. 

We start with an initial build of the codebase, which creates the initial versions \smalltt{M.V0},
\smalltt{B0.V0}, and \smalltt{B1.V0}. 
In step~(1), the user optimizes method \smalltt{M} and rebuild the codebase, which creates the version
\smalltt{M.V1}.
% As \smalltt{M.V1} replaces the previous version~\smalltt{M.V0}, the former directly succeeds the latter in the version graph.
Since the other methods were not modified and also did not undergo an anomaly, their initial versions
and results remain valid, which is indicated by the small black circles in the graph. 
In step~(2), the users optimizes~\smalltt{M} again and recompiles, resulting in the compilation of a new
version~\smalltt{M.V2}. However, the generated code of \smalltt{B0} changed over the previous build,
which \m{} detects and classifies as an anomaly, since the corresponding source code has not been
modified. In the version graph, this anomaly is reflected by a fork to version~\smalltt{B0.V1}.
Due to the fork, both versions \smalltt{B0.V0} and \smalltt{B0.V1} would be considered for result
interpretation.
In step~(3), let us now assume that \smalltt{M} is modified once more, which now causes two
side-effects: First, it triggers an anomaly in \smalltt{B1}, resulting in a fork between
\smalltt{B1.V0} and \smalltt{B1.V1}. Second, the previously occurred anomaly in \smalltt{B0}
disappears and \smalltt{B0} compiles again to the previously seen version \smalltt{B0.V0}.
Both versions of \smalltt{B0} still need to be considered for result interpretation.
% At this point both versions of \smalltt{B0} have to be considered for result interpretation.
% At this point both versions \smalltt{B0.V0} and \smalltt{B0.V1} have to be considered for result interpretation.
% This must \textit{not} be reflected in the version graph, as both version should still be considered during result interpretation.
%
This changes in step~(4), when the users actively modifies the baseline~\smalltt{B0}, e.g., to fix a bug.
This modification obviously creates a new version \smalltt{B0.V2}, which merges the fork, indicating
that the previous versions \smalltt{B0.V0} and \smalltt{B0.V1} are outdated and
should no longer be considered.
In summary, at any point in time, \m{} locates for each method all relevant versions by traversing from
top to bottom along all paths until the first version on each path is visited. These versions in
conjunction are considered during result interpretation. In Figure~\ref{fig:version_graph}, we mark
all versions that are relevant after the last step in red boxes.

% reset boundary, when the environment changes (workload, dataset etc)

\section{Architecture and Implementation}

\begin{figure}[h!]
    \vspace*{-0.3cm}
    \centering
    \includegraphics[page=3, width=.97\columnwidth, trim={0 5.8cm 5.3cm 0}, clip]{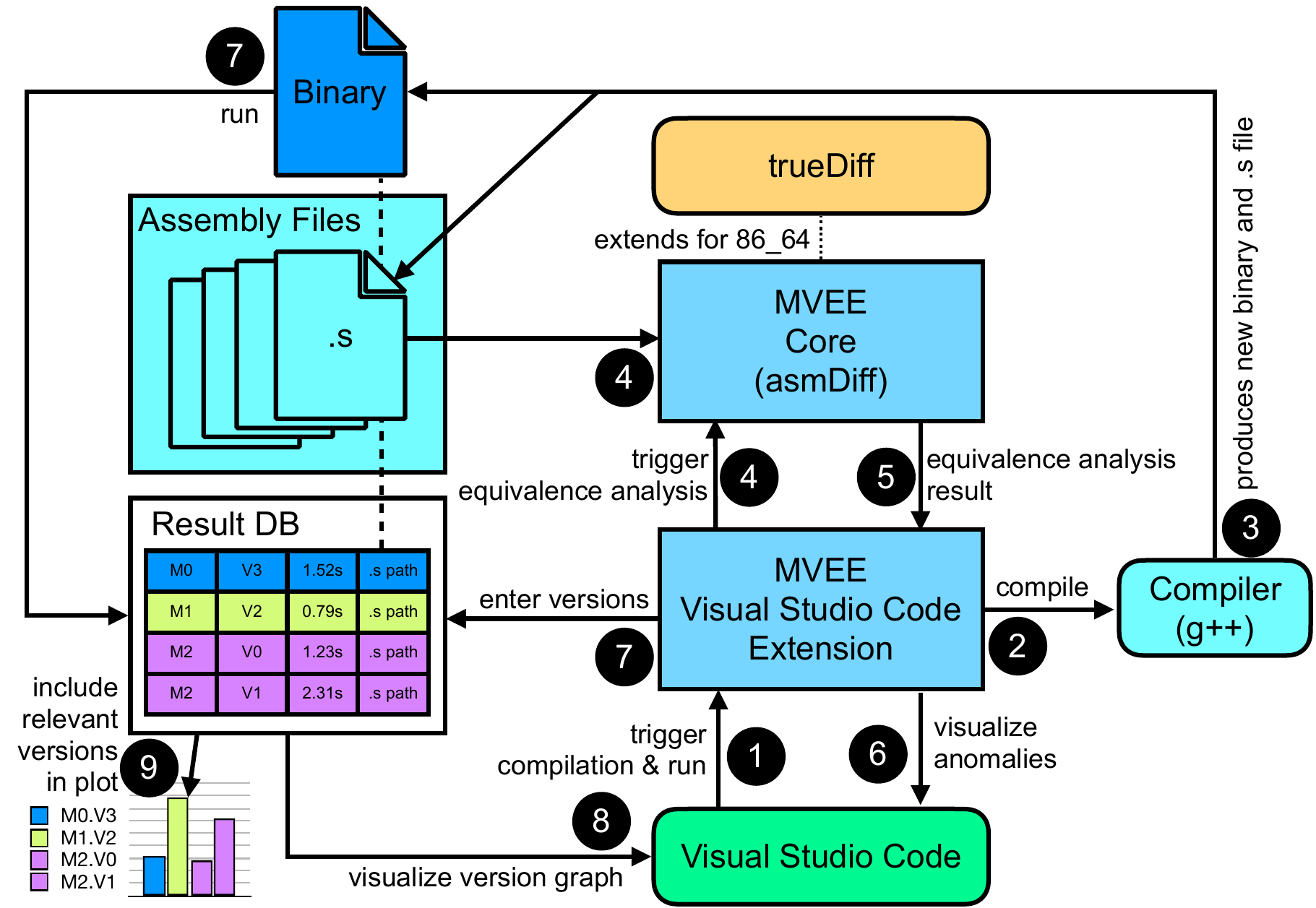}
    \vspace*{-0.2cm}
    \caption{High-level architecture and interaction.}
    \label{fig:architecture}
    \vspace*{-0.4cm}
\end{figure}

Figure~\ref{fig:architecture} provides a high-level view on the components of our \m{} implementation and how they interact with each other. \m{} itself is split into two components: The VSCode extension and \mc{}, which handles the assembly file analysis.

\begin{figure*}[ht!]
    \centering
    % \includegraphics[width=.97\linewidth]{figures/screenshot.png}
    % \includepdf[width=.97\linewidth]{figures/annotated_screenshot_dark.pdf}
    \includegraphics[width=.995\linewidth, trim={0 0cm 0 0}, clip]{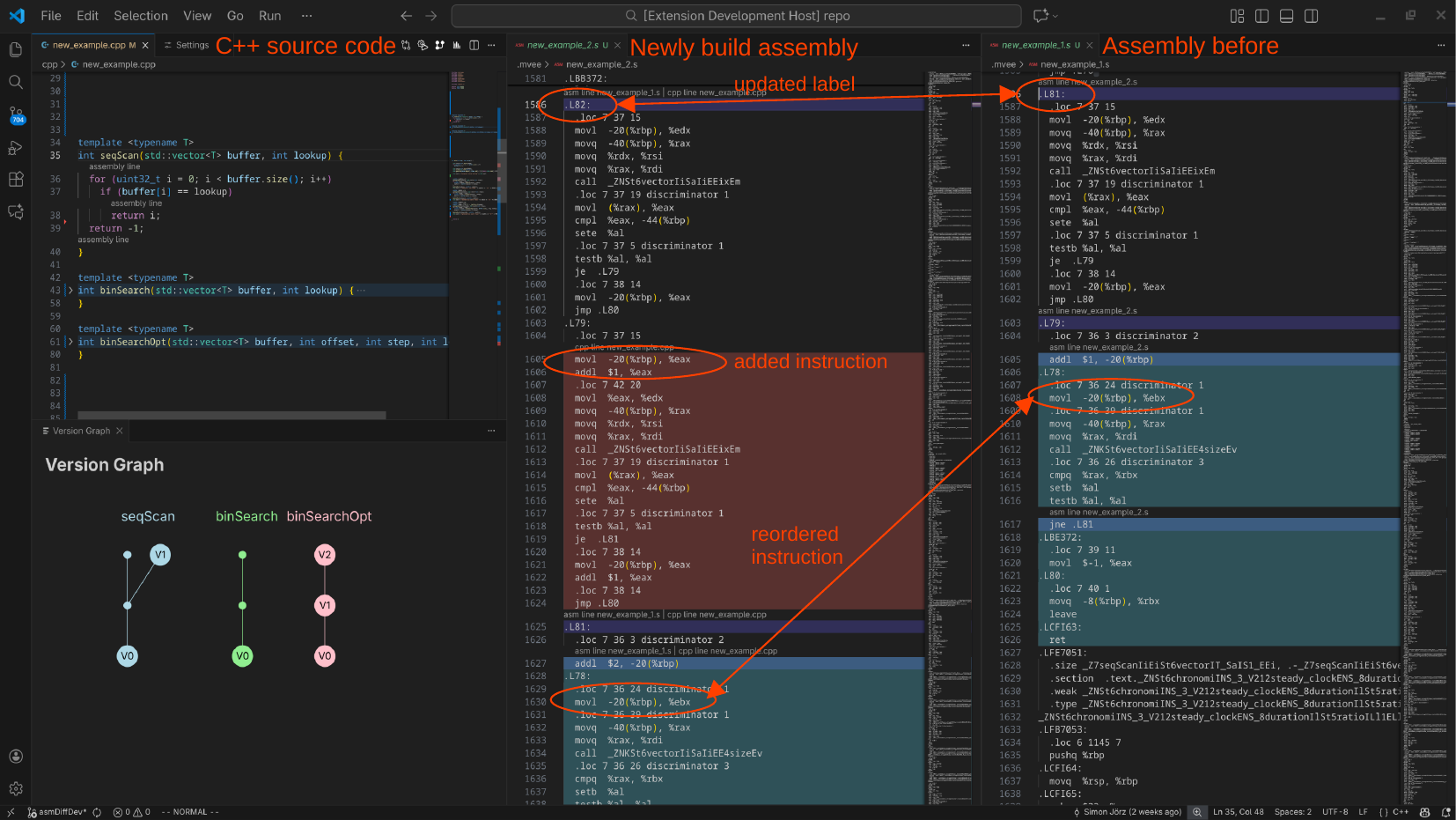}
    \vspace*{-0.4cm}
    \caption{The inspection of a detected anomaly in our \m{} VSCode extension.}
    \label{fig:screenshot}
    \vspace*{-0.5cm}
\end{figure*}

\m{} operates as follows: In~\circled{1}, from the IDE, the user tells the \m{} extension to compile the current codebase in order to perform a new experimental run. In \circled{2}, our extension then issues the compilation, which in \circled{3} produces a new binary and the corresponding \smalltt{.s} file. This assembly file will be stored with the previous builds in the assembly files directory, named by the timestamp of the run. Next, in~\circled{4}, our extension instructs \mc{} to perform an equivalence analysis for all methods that were not modified in the source code since the last build. This happens based on the corresponding \smalltt{.s} files. The core component passes the result of this equivalence analysis for each compared method in \circled{5} to our extension --- if anomalies have been detected, they can be inspected by the user in \circled{6} directly in the IDE. In \circled{7}, the extension now triggers the actual run of the binary, which produces a new result for each method. If an anomaly has been detected before in a method, the new result of that method enters the result database as a new version. Otherwise, the current version is simply replaced by the new version.
Additionally, in \circled{8}, the version graph that is displayed in the IDE is updated accordingly. Finally, in \circled{9}, all relevant versions of the result database are extracted from the result database to generate a corresponding results plot.

\vspace*{-0.2cm}
\subsection{Assembly Code Analysis}
\label{ssec:truediff}
\vspace*{-0.1cm}

% General workflow

The actual assembly code analysis operates as follows: Given a pair of assembly files to
compare, \mc{} first parses each file and computes a corresponding tree-like intermediate
representation for the regions of code that are relevant. Then, for the comparison, \mc{} 
leverages \textit{truediff}~\cite{lit:truediff}, a tree based structural diffing algorithm which is 
designed to return concise and type safe edit scripts for typed tree-shaped data. To be usable in
our context, we extended \textit{truediff} to handle \smalltt{x86} assembly code. As a result of the
diffing, \textit{truediff} produces a so-called \textit{edit script}, which captures all the
observed differences between the analyzed assembly files in a format called \textit{truechange}.
\mc{} then goes through the edit script and identifies all differences that are relevant for our
definition of equivalence (see Section~\ref{ssec:mvee_intro}). In the following, we outline the
steps in detail:

% Detailed workflow
% Part 1: From the assembly files to the truediff input

\textbf{Assembly files $\rightarrow$ \textit{truediff} input}. \mc{} first parses all assembly
instructions of a file and generates a corresponding intermediate representation. This
representation is kept general to be able to represent the large                      % TODO: this way we it is quite simple to adjust to other dialects
backwards-compatible \smalltt{x86} instruction set without having to specify
each and every instruction individually. From the intermediate representation, \mc{} next extracts
only the code portions that are relevant for the experimental evaluation, which relies on the marks
placed in the C++ code.
Precisely, \mc{} indirectly builds and traverses the control flow graph from each start
mark until the corresponding end mark.
% TODO: explain grouping into Fall Through Groups
% After extracting all basic-blocks that contain relevant code lines, \m{}groups blocks that are
% reachable through a fallthrough.
% The control flow can exit a basic-block in two different ways. Either by a control altering instruction
% (transferring the control to any basic-block in the program), or by ``falling through'' into the next basic-bock.
% This ``fallthrough'' happens simply by the lack of a control altering instruction, where the program executes
% the next line, which happens to be the first line in the next basic-block.
To keep this control flow intact, we group instructions into fallthrough-groups.
Each fallthrough-group is a sequence of instructions where one can be executed after another without
an unconditional jump or call occurring in-between. Such fallthrough-groups can be reordered
without changing the flow of control.
% Note, that the control can still move from one fallthrough-group into another by jumps or calls to labels.
% This helps later to analyze whether the control flow of such a fallthrough breaks.
% Further, \mc{} groups basic-blocks into so-called
% fallthrough-groups. Each fallthrough-group is a sequence of basic-blocks where a block can reach all
% other blocks further down in the sequence by ``falling through'' the others. In the later analysis of the generated edit scripts, this allows to analyze whether an edit breaks such a sequence of fallthroughs.

\textbf{Fallthrough-groups $\rightarrow$ edit script.} Next, \mc{} passes the fallthrough-groups of
both assembly files as source tree and target tree to the \textit{truediff} algorithm to compute the
edit script. This edit script describes how to modify the source to obtain the target.
On a high level, \textit{truediff} encodes the structure and literals of all contained subtrees as
hashes and identifies similarities by comparing these hashes. Subtrees of the source that match with
subtrees of the target are reused, while non-matching subtrees are deleted or inserted as needed.
% For matched subtrees of the source tree,
% edits are chosen in such a way that they are moved and updated to equal their matched subtree in the 
% target tree. Subtrees of the source tree that are not matched are removed completely, while subtrees
% of the target without a match are inserted.
In this way the \textit{truediff} algorithm creates a concise edit script, while ensuring to use
every node exactly once, within linear run-time complexity.

% The resulting edit script can consist of the following three categories of edits: 
% (a)~\update{}, which represents updates to literal values.
% (b)~\detach{} and \attach{}, which represent a detach respectively attach of a subtree from the tree based structure.
% (c)~\unload{} and \load{}, which represents a removal respectively creation of a subtree.

% To match similar or equal parts of a graph \textit{truediff} uses two cryptographic hashes, namely
% the structural hash and the literal hash. The structural hash represents the structure of a node and
% all its children. This hash excludes any information about actual values. In contrast, the literal
% hash stores that literal information for each node in the subgraph.
% Therefore, if two subgraphs are equal, their structural and literal hashes are as well.
% \textit{truediff}'s equality check operates on a source and target tree as follows: The algorithm first computes
% both the structural and literal hash for every subgraph in both trees. Then, it traverses the source
% tree and groups subgraphs with the same structural hash. Afterward it traverses the target tree
% assigning each subgraph to exactly one of the source tree with the same structural hash. If there is
% a subgraph that also has the same literal hash and therefore is equal, it is preferred as a match.
% During this process, matched elements are removed from their group allowing exactly one match per subgraph.
% After this assignment is done, the actual edit script is computed. 

\textbf{Edit script $\rightarrow$ anomalies}. Then, \mc{} analyzes whether the edit script violates
our definition of equivalence in three steps:
% 1.
(1)~\mc{} checks whether the script removes or inserts any new instructions, operands or labels. 
The existence of such significant structural edits clearly violates equivalency.
% 2.
(2)~\mc{} checks whether the updates do not change any data and are consistent.
Precisely, it gathers updates to operands (memory references or immediate values). If updates change an 
immediate or a memory reference this is considered as a violation of the equivalency. If an update
changes a label, \mc{} verifies that all other references to that label are changed as well to the
same name. Is that not the case, then the code is as well considered to be not equivalent.
% 3.
(3)~\mc{} checks whether the reordering of code changes the control flow.
Reordering a fallthrough-group does not change the control flow, as jumps remain consistent. But
reordering single instructions (or operands) does change the control flow. In the latter case \mc{}
consideres a reordering as a violation of the equivalency.
Finally, it returns the equivalence result and all edits.

\section{User Experience}

Figure~\ref{fig:screenshot} shows how the equivalence result is presented to the user in our VSCode extension. Using \smalltt{CodeLens}, our extension links the detected anomaly to the corresponding C++ source code lines, such that the user sees from where the anomaly originates. By clicking on the link, the corresponding assembly file opens, in which all edits are highlighted in different colors depending on their category. 
Additionally, the \m{} extension continuously visualizes the version graph in a git-style manner.

In the demonstration, the audience is able to experience the behavior of the extension and the benefits of \m{} at the example of multiple observed real-world anomalies which have been produced during compilation with the general-purpose compiler~\smalltt{gcc}. These real-world anomalies include the discussed case from our motivating example~\cite{lit:compilation_anomalies}, but also further observed anomalies --- identified by colleagues in their own benchmarking code bases but also found via automated extensive search using AI~tools. 
To ensure that the audience actually experiences anomalies on site, we prepare for all code bases in advance a set of snapshots that capture the state between critical source code changes that lead to the generation of compilation and performance anomalies. The audience can then (a)~observe how our MVEE extension automatically detects the anomalies in the respective code parts, and how they are correctly registered as new versions in the version graph. Additionally, (b)~the audience is able to analyze the type of anomalies both on the C++ and assembly level in detail, as shown in Figure~\ref{fig:screenshot}. To experience the impact on an experimental evaluation, the audience is further able to (c)~produce runtime plots that correctly include all identified versions that are relevant for the experimental evaluation.

% At the venue, we will present our \m{} extension running in VSCode as well as a benchmarking
% codebase that compare performance-critical methods.
% The audience will be able to choose between a selection of snap-shots. Each snap-shot consists of
% a codebase right after an anomaly happened.
% The audience can then inspect the C++ source code and version history. From the source code they
% will be able view the assembly code and visualize the difference between different versions. Furthermore they will
% be able to plot the run time of each of the compilations. To get a feeling for the versioning, the
% audience will also be able to produce new results. They can change the source code, compile it and
% view how the versioning happens in the background.
%
% In order to provide interesting real world examples of anomalies we will make a thorough search.
% On the one hand, we want to ask the community for cases where they suspect that such anomalies could
% have happened in the past. This includes the anomaly presented in \cite{lit:compilation_anomalies}.
% On the other hand, we want to make an extensive automated search on widely used web platforms such as GitHub.
% %
% This should raise awareness of this important problem and present a possible solution on how to handle it.


\vspace*{-0.2cm}

\bibliographystyle{ACM-Reference-Format}
\bibliography{acmart}

\end{document}